\documentclass[twocolumn,showpacs,preprintnumbers,amsmath,amssymb,letterpaper]{revtex4}
\usepackage{amssymb}
\usepackage{graphicx}% Include figure files
\usepackage{dcolumn}% Align table columns on decimal point
\usepackage{bm}% bold math

\begin{document}

\title{Experimental and theoretical study of processes
of formation and growth of pearlite colonies in eutectoid steels}
\renewcommand{\abstractname}{}
\author{ V. G. Vaks$^{a,b}$,  A. Yu. Stroev$^{a,b}$,  V. N. Urtsev$^c$ and A. V. Shmakov$^c$ }
\affiliation{$^a$Russian Research Center "Kurchatov Institute",
123182
Moscow, Russia\\
$^{b}$Moscow Institute of Physics and Technology (State University),
117303 Moscow, Russia\\
$^c$Research-Technological Centre ``Ausferr'', 455023, Magnitogorsk,
Russia}

\date{\today}

\begin{abstract}
We describe our optical and electron-microscopy observations of
pearlite structures in eutectoid steels which seem to imply that the
mechanisms of formation of pearlite colonies in these steels differ
from those  observed earlier for non-eutectoid steels. A simple
theoretical model to study kinetics of pearlite transformations is
suggested. Simulations of growth of pearlite colonies based on this
model reveal that for the volume carbon diffusion mechanism
usually-supposed such growth is always unstable, and the
steady-state growth can be realized only via the interfacial carbon
diffusion mechanism. A model of formation of pearlite colonies based
on the assumption of a strong enhancement of carbon diffusion near
grain boundaries is also suggested. The model can be applicable to
the plastically deformed steels, and the results of simulations
based on this model qualitatively agree with some microstructural
features of formation of pearlite colonies observed in such steels.
\end{abstract}


\maketitle

\section{ INTRODUCTION}

Studies of growth and formation of pearlite colonies in steels
attract great attention for many decades, see, e. g.
\cite{Mehl-56}-\cite{Steinbach-07} and references therein. However,
both the experimental information and  the theoretical understanding
of these phenomena seem to remain to be rather limited. Most of
experimental data used in the literature (see e. g.
\cite{Mehl-56,Hillert-62}, \cite{Whiting-00}-\cite{Steinbach-07})
were obtained long ago and by not modern methods, and many important
phenomena, in particular, formation of pearlite colonies, seem to be
insufficiently studied. In theoretical treatments, only steady-state
growth of colonies is usually considered, basing mainly on the
equilibrium thermodynamics ideas \cite{Zener-45}-\cite{Hillert-72},
but with no attempts of microscopic treatments of kinetic processes.
Recently, phenomenological phase-field approaches were used to treat
the pearlite growth problem \cite{Steinbach-06,Steinbach-07}.
However, these treatments  employ many phenomenological parameters
whose physical meaning is not always clear (while no microscopic
estimates for them are given), and they discuss only volume but not
interfacial diffusion mechanisms (which can be very important for
such processes, see \cite{Plapp-09} and below). Thus the results
obtained can hardly significantly elucidate the actual microscopic
mechanisms of growth of pearlite colonies.

 At the same time, even main mechanisms of pearlite transformations seem to be
not well understood as yet. It is unknown whether the colony growth
kinetics is determined by the volume or the interfacial diffusion of
carbon atoms, and this problem is debated
\cite{Whiting-00,Plapp-09}. It is not clear which kinetic or
thermodynamic factors determine the colony period \,$S$\,
\cite{Mehl-56,Schastlivtsev-06}. There are absent not only theories
but even any definite ideas about the mechanism of formation of
colonies near grain boundaries and other lattice defects
\cite{Mehl-56,Hillert-62,Schastlivtsev-06}, etc.

This work aims to study some of the problems mentioned both
experimentally and theoretically. In sec. 2 we present some results
of our experimental observations of processes of formation of
pearlite colonies in an eutectoid steel. These processes are widely
discussed in the literature. It is generally believed that the
colonies are formed near grain boundaries of austenite or near other
defects of crystal lattice \cite{Mehl-56,Hillert-62}, but the
detailed experimental information about these processes is rather
scarce as yet. Most thoroughly they have been discussed by Hillert
\cite{Hillert-62} who studied formation of pearlite colonies in both
hypoeutectoid steels (those with the carbon concentration \,$c$\,
lower than its eutectoid value \,$c_e$)\, and hypereutectoid steels
with \,$c>c_e$.\, Hillert found that the earlier-supposed ``repeated
sidewise growth'' mechanism \cite{Mehl-56} is not observed in his
experiments; instead, ``the individual lamellae form by branching
during edgewise growth'' near various lattice defects: grain
boundaries, twinned planes, interphase boundaries, etc. However, the
eutectoid steels with \,$c\simeq c_e$\, were not studied in Ref.
\cite{Hillert-62}. Therefore, it seems interesting to study the
colony formation processes in such steels by the modern methods. In
Sec. 2 we describe such studies and compare the results of our
observations to those of previous work \cite{Mehl-56,Hillert-62}.

The rest part of this work describes some attempts to develop the
microscopic approach to  the pearlite transformation theory. As a
possible first step for that, a simple model of alloys iron-carbon
was recently suggested  \cite{VS-08}. This is a binary interstitial
alloy model with the phase diagrams illustrated by Fig. 5(a) below
being symmetric with respect to the interchange of ferrite and
cementite. The pearlite growth kinetics  for this model was studied
basing on the appropriate Ginzburg-Landau functional. The  volume
diffusion of carbon was supposed to be the main kinetic mechanism of
this growth, in accordance with the most of models used in the
literature \cite{Mehl-56}-\cite{Hillert-72}. The results of
simulations of growth of colonies made in \cite{VS-08} agreed
basically with the conclusions of phenomenological treatments
\cite{Zener-45}-\cite{Hillert-72}, though some new effects have also
been found.

However,   the real phase diagram ferrite-cementite (if we model it
by a binary alloy model) is by no means symmetric with respect to
the interchanges of ferrite and cementite. For example, if we take
the carbon concentration in cementite for unity, then the eutectoid
concentration is actually close to \,$c_e$=$1/8$,\, instead of
\,$c_e$=$1/2$\, in symmetric models. Therefore, to make such
approach more realistic, we should study the colony growth also for
non-symmetric models with \,$c_e\simeq 1/8$.\, This is done in Secs.
3 and 4 of the present work. We find that for all models of the type
considered, the growth of colonies via the volume carbon diffusion
mechanism is unstable, and the steady-state growth can be realized
only via the interfacial diffusion of carbon. It agrees with the
similar conclusion of a recent empirical analysis \cite{Whiting-00}.

In Sec. 5 we suggest a model of formation of pearlite colonies near
grain boundaries of austenite. As mentioned, the colonies are
usually formed  just in this region, but until now there seem to be
no microscopic explanations for that \cite{Mehl-56,Hillert-62}. In
the model suggested, we relate the formation of colonies to a great
enhancement of carbon diffusion  near grain boundaries discussed by
a number of authors \cite{Bokshtein-61,Kesarev-10}. The results of
our treatment show that in the simplest form presented this model
can be applied only to the strongly deformed materials, such as
those studied in Ref. \cite{Tushinsky-93}. However, further
generalizations can extend the region of applicability of the
approach suggested. Main conclusions are summarized in Sec. 6.

\section{Observations of microstructure of pearlite colonies in
non-deformed and deformed eutectoid steels}

\subsection{Preparation of samples and methods of experiments\label{samples}}

All microstructural investigations were made on samples of steel
\,85\, with the composition shown in Table 1. Thus, the
concentration of carbon was close to the eutectoid one for alloys
iron-carbon: \,$c_e^{\rm Fe-C}\simeq 3.45\%$.\, It differs the steel
used in our study from the hypoeutectoid steels with \,$c<c_e$\, and
hypereutectoid steels with \,$c>c_e$\, investigated by Hillert
\cite{Hillert-62}.

 \vskip5mm
\noindent {\bf Table 1.} Composition of samples.
 \vskip3mm
{\begin{tabular}{|c|c|c|c|c|c|c|c|c|c|} \hline
 &&&&&&&&&\cr
Impurity&C&Mn&Si&Cr&Ni&Cu&S&P&Al\\
\hline &&&&&&&&&\cr
at. \% &3.91&0.60&0.48&0.08&0.08&0.04&0.04&0.02&0.02 \\
\hline
\end{tabular}
 \vskip5mm

Rods of steel 85 of diameter 8 mm and length about 500 mm were
subjected to various thermal and mechanical treatments described
below. These different kinds of treatment are numbered by the same
way as figures 1--4 in which we show the microstructure of samples
for each kind of treatment.

(1) Heating in a muffled furnace up to 1050$^o$C, cooling at air up
to 650$^o$C, putting sample in a furnace heated up to 650$^o$C,
annealing at this temperature for 20 sec, then quenching in water.

(2)  Heating in a muffled furnace up to 1000$^o$C, cooling in the
melt of sodium nitrate NaNO$_3$  at temperature  443$^o$C up to
reaching the melt temperature,  then quenching in water.

(3)  Heating in a muffled furnace up to 1005$^o$C, cooling at air up
to 750$^o$C , plastic deformation at this temperature in a torsional
type plastometer with the deformation rate 0.2/sec up to the strain
23\% (for the surface layer), cooling in the melt of  NaNO$_3$  at
temperature 443$^o$C up to reaching the melt temperature, then
quenching in water.

(4)  Heating in a muffled furnace up to 1000$^o$C , cooling in the
melt of NaNO$_3$  at temperature  443$^o$C  up to temperature
750$^o$C , then quenching from 750$^o$C   in water.

Samples for  microstructural studies were cut in the shape of thin
washers parallel to the cross-section of the rod. Then they were
subjected to a many-stage grinding and polishing procedure followed
by etching in a 5\% solution of nitrogen acid in ethyl alcohol
(nital).

For a sample treated according to regime (1), the photograph shown
in Fig. 1  has been obtained using the structure analyzer Nikon
Epiphot-TME and camera Panasonic WV-CP610/G.

Microstructure of samples  treated according to regimes (2)-(4) was
studied by the scanning electron microscopy (SEM) methods. Cuts
prepared for the optical microscopy studies were used for SEM, too.
However, these cuts were subjected to the additional ion etching. It
provides a high cleanness of the surface and enables one to reveal
fine details of the structure \cite{Bokshtein-04}. The high
resolution scanning electron microscope "SUPRA 55" by LEO (ZEISS)
with ``Gemini'' field emission column was used to acquire the images
shown in figures 2--4. The in-lens secondary electron detector
combined with a low accelerating voltage (1.5 keV) was found to
provide the best quality images. Given that the plates of pearlite
intersect the surface of the cut at random angles, no sample tilt
was required.

The SEM method was chosen because its high resolution makes it to be
particularly suitable for the metallographic studies of alloys with
a dispersed structure, such as the eutectoid mixtures treated in
this study. The characteristic feature of a topographic contrast in
SEM is the enhanced brightness of images of sharp hills and ledges
of a surface relief, the so-called ``edge-effect'' related to the
increased output of electrons from such areas \cite{Bokshtein-04}.
This enhanced brightness is seen in figures 2--4. The surface of
pearlite areas in these figures looks as a well-developed relief
formed by the combined action of the mechanical loading under
polishing and the subsequent etching by nital \cite{Hillert-62}. At
the same time, martensite (obtained from the untransformed
austenite) is  by about an order of magnitude harder than ferrite,
and it does not contain cementite inclusions. Therefore, martensite
areas at the cut surface are much more flat and uniform than those
of ferrite and pearlite, and these martensite areas are seen in
figures 4(a) and 4(b) as a grey background.

\subsection{Experimental results\label{exp-results}}

It is generally accepted that ``in the homogeneous austenite,
pearlite is formed practically only on grain boundaries''
\cite{Mehl-56}. Our observations agree with this conclusion. In Fig.
1 we show the network of decarburizing formed at the surface of a
sample. This network is formed due to the chemical interaction of
carbon contained in the steel with the atmosphere oxygen, which
results in the formation of gaseous oxides and thus in the removal
of carbon from the steel. In Fig. 1, the original grain boundaries
of austenite are decorated by chains of tiny bright grains of
ferrite. Such decoration arises because the surface decarburizing
takes place primarily on the grain boundaries. It is so because,
first, the velocity of grain-boundary diffusion of carbon much
exceeds that of its diffusion within the grain and, second, because
the grain-boundary regions are much easier to access for oxygen as
compared to inner parts of grain. Fig. 1 also shows that the
positions of these ferrite grains seem to coincide with the probable
initial positions of pearlite colonies (seen as dark areas), thus
the colonies seem to spread just from these chains. This observation
agrees with the above-mentioned conclusion that formation of
pearlite colonies takes place on the original grain boundaries of
austenite.

Fig. 2 shows the microstructure of a classical tiny-dispersed
plate-like pearlite, or sorbite, for which inter-plate distances, or
colony periods \,$S$,\, are about 100-200 nm. The well-developed
relief consisting of relatively thin cementite plates in the ferrite
matrix is formed due to the many-stage surface treatment (polishing
and etching) described in Sec. 2.1 which removes the surface layer
of ferrite but make little effect on cementite. Therefore, dark
strips (``cavities'') in Fig. 2 correspond to the original (and
removed) ferrite, while bright areas (``ridges'' of different tilt
to the cut plane),  correspond to cementite.

It seems natural to suggest that the broad approximately vertical
region of ferrite in the middle of Fig. 2, from one side, and the
thin layer of cementite along the left edge of this region together
with a lot of approximately horizontal colonies adjacent to this
layer, from other side, correspond to two opposite sides of an
original grain boundary of austenite. Then the enrichment of this
left region by carbon (being necessary, in particular,  to form the
boundary cementite layer mentioned) is naturally explained by
``releasing'' this carbon from ferrite regions formed to the right
of this grain boundary. Therefore, it seems most probable that the
above-mentioned approximately horizontal colonies to the left of
this boundary have been formed just on the boundary cementite layer
(or near it). At the same time,  for pearlite colonies to the right
of this boundary, it seems natural to suggest that they have been
formed from the ferrite layer. Therefore, Fig. 2 can illustrate both
two earlier-discussed mechanisms of formation of pearlite colonies
\cite{Mehl-56,Hillert-62}, that from the primary cementite and that
from the primary ferrite, as well as the probable close
interrelation of these two mechanisms in the eutectoid steel under
consideration.

One can also observe in Fig. 2  a number of ``defect'' lines (that
is, cuts of appropriate surfaces) where different colonies seem to
meet each other or overcome some  obstacles. Most of these lines
seem to correspond just to ``collisions'' of colonies growing from
different sides of a grain, but some of them can also correspond to
lattice defects, for example, to twin boundaries or subgrains of
original austenite.

\begin{figure}
\caption{Network of decarburizing along original grain boundaries of
austenite observed for a sample prepared as described in point (1)
of Sec. 2.1.  Dark areas correspond to pearlite, bright inclusions
within these region, to ferrite grains, and the grey matrix, to
martensite formed after a quench of an untransformed
austenite.\label{grain-boundaries}}
\end{figure}

\begin{figure}
\caption{Microphotograph of a nondeformed sample  prepared as
described in point (2) of Sec. 2.1. \label{pearlite-non-deformed}}
\end{figure}
\begin{figure}
\caption{Microphotograph of a deformed sample  prepared as described
in point (3) of Sec. 2.1.  Frames (a), (b) and (c) correspond to
different areas of a sample. \label{pearlite-deformed}}
\end{figure}

In the plastically deformed samples, both the transformation
kinetics and microstructure change considerably. The transformation
rate significantly increases. In our experiments, after the hot
treatment described in point (3)  of Sec. 2.1, the full
transformation time for a cylindric sample of 8 mm length decreased
from 10 to 4 seconds. The degree of order and parallelism of
pearlite colonies are notably reduced, and many colonies are
fragmented, which is illustrated by frames  3(a)-3(c). However, the
main microstructural features of formation of colonies discussed
above seem to be present in these samples, too. For example,  in
frame 3(a) one can distinctly see the cementite layer separating
ferrite and plate-like pearlite regions, and this layer (together
with its vicinity) again seems to be the most probable place of
formation of plate-like pearlite colonies. Analogous places of a
probable formation of plate-like pearlite colonies (being, though,
less regular) are seen in frames  3(b) and 3(c). At the same time,
on ``ferrite'' sides of original grain boundaries, the formation of
plate-like pearlite colonies  in deformed samples seems to be
hampered: according to frames 3(a)-3(c), cementite precipitates in
this regions have usually an irregular shape being more close to the
globular than to the plate-like one.

\begin{figure}
\caption{Microphotograph of a sample  prepared as described in point
(4) of Sec. 2.1.  Frames (a) and (b) correspond to different areas
of a sample. \label{pearlite-deformed-start}}
\end{figure}

In one of experiments with non-deformed samples, varying cooling and
temperature conditions, we succeeded to fix an initial stage of
pearlite transformation. Microstructures observed in these
experiments are illustrated by Fig. 4 where we see some transformed
regions surrounded by the region of martensite (grey colored) formed
under a quench of an untransformed austenite. More dark areas again
correspond to the original ferrite removed by the polishing and
etching treatments mentioned. In frame 4(a), in the part of the
transformation volume adjacent to its upper boundary, we seem to
observe both formation of plate-like pearlite colonies ``from
ferrite'' (via the mechanism similar to that shown in Fig. 2 to the
right of the original grain boundary), and non-plate-like,
``globular-type'' precipitates of cementite similar to those seen in
Fig. 3. At the same time, near right boundaries of the transformed
region we see rather regular plate-like colonies (tilted to the cut
plane) which could be formed via the mechanism ``from the boundary
cementite layer'' similar to that discussed for the left part of
Fig. 2. In frame 4(b), analogous rather regular colonies are
observed near lower right boundaries of the transformed region, thus
vicinities of these boundaries can be the regions of formation of
pearlite colonies, too.

\subsection{Discussion of experimental results\label{discussion-exp-results}}

The above-described observations enable us to make following
conclusions and conjectures about mechanisms of formation of
pearlite colonies.

1. These observations agree with the point of view generally
accepted \cite{Mehl-56} that  in the homogeneous austenite, the
plate-like pearlite colonies form mainly near its grain boundaries.

2. At the same time, our observations seem to imply that the
mechanism of this formation is complex and includes at least two
different stages. The first stage seems to correspond to the
decomposition of austenite into ferrite and cementite within some
immediate vicinity of the grain boundary accompanied by a transfer
of carbon through this boundary. On the side of this boundary
enhanced by carbon, the above-mentioned cementite layer is usually
formed, and the ferrite layer is formed on the opposite side of this
grain boundary.

3. Then the plate-like colonies start to form, being usually
approximately normal to the original grain boundary. The  mechanisms
of this formation are not clear as yet. However, these mechanisms
seem to be certainly different for the ``cementite'' and ``ferrite''
sides of this boundary, that is,  for the regions enriched and
depleted in carbon. It seems to follow, in particular, from a quite
different morphology of new-formed colonies in these two regions
illustrated by Fig. 2. These differences become still more evident
for the plastically deformed samples when the plate-like structure
of pearlite colonies usually remains only on the ``cementite'' side
of the boundary, while on its ``ferrite'' side, precipitates of
cementite have often an approximately globular rather than the
plate-like structure.

4. The above-mentioned conclusions about the ``many-step'' character
of formation of pearlite colonies agree qualitatively with the
similar conclusions made by Hillert \cite{Hillert-62}. However, he
studied the hypoeutectoid and hypereutectoid steels rather than the
eutectoid ones considered  in the present work. In addition to that,
Hillert studied mainly processes of formation of pearlite due to the
interaction of initial lamellas of ferrite or cementite with various
lattice defects, such as twin boundaries, grain boundaries,
cementite layers on grain boundaries, etc. The main mechanism of
formation of colonies observed in his studies  was ``branching'' of
this initial lamellas. For the formation of colonies near grain
boundaries of the homogeneous eutectoid austenite studied in our
work, such branchings have not been observed. Therefore, one can
believe that these branchings are characteristic of formation of
colonies in  more complex conditions considered by Hillert
\cite{Hillert-62} while we observe more simple processes.

\section{Microscopic models for description of  pearlite transformations}

As mentioned, even main mechanisms of  pearlite transformation seem
to be not quite clear as yet.  Therefore, in theoretical approaches
to this problem it seems desirable to use only simplest models which
yield a reasonable description of phase equilibria between
austenite, ferrite and cementite, while for the rest include as
small number of model parameters  with a clear physical meaning as
possible. In the previous work \cite{VS-08}, such model has been
suggested  to study the colony growth in alloys with symmetrical
phase diagrams of the type shown in Fig. 5(a). Below we generalize
this model to the case of more realistic, non-symmetrical phase
diagrams.

\subsection{Thermodynamic model\label{thermodynamic-model}}

We consider the interstitial alloys  \,\hbox{MeX$_c$} in which the
number of interstices (interstitial voids) for atoms \,X\, is equal
to the number of metallic atoms \,Me,\, as in austenite and in
simplified models of cementite \cite{VKh-2}. Then the concentration
\,$x$\, of interstitial atoms \,X\, is related to the average
filling of interstices  \,$c$\, by the relation: \,$c=x/(1-x)$,\,
but for brevity, the term ``concentration'' will be used below for
the average filling of interstices, \,$c$.\, We consider the models
with equilibrium phase diagrams temperature $T$ - concentration
\,$c$,\, shown in Fig.  5, in which phases  $A$, $D$ and $B$ are
analogues of ferrite, austenite and cementite, respectively. Phase
$D$ is treated as a disordered solid solution of interstitial
(carbon) atoms in the FCC lattice described by the mean-field
(``regular solution'') approximation, while the phases $A$ and $B$
are described by the order parameters  \,$\eta$\, и \,$\zeta$\,
which for the uniform equilibrium phases have the following values:

\begin{eqnarray}
&&A:\ (\eta =1,\zeta =0),\qquad  B:\ (\eta =0,\zeta =1),\nonumber\\
&& D:\ (\eta =0,\zeta =0).\label{eta-zeta}
\end{eqnarray}
For the iron-carbon  alloys, the  parameter\,$\eta$\,  can be
considered to be proportional to the Bain or Kurdyumov-Zaks
deformation realizing the FCC-BCC transformation, and \,$\zeta$\, is
an analogous parameter describing the transition from austenite to
cementite \cite{VKh-2}.

The inhomogeneous alloy states under consideration are described
using the generalized Ginzburg-Landau functional  \,$F$\,
\cite{PV-03,VKh-5} supposing that the characteristic inhomogeneity
lengths \,$l_{inh}$\,  in the \,$c({\bf r})$,\, \,$\eta({\bf r})$\,
and \,$\zeta({\bf r})$\, coordinate dependences much exceed the
lattice constant \,$a$:\,
\begin{equation}
F=\int \frac{d^3r}{v_a}\left[ f(c,\eta,\zeta)+G(\nabla c, \nabla\eta
,\nabla\zeta)\right]. \label{F_GL-tot}
\end{equation}
Here \,$v_a$\, is volume per iron atom,\, and \,$f(c,\eta,\zeta)$\,
is the free energy of a uniform alloy per iron atom. The gradient
term \,$G$\, is a bilinear form of  gradients \,$\nabla c$,\,
$\nabla\eta$\, and \,$\nabla\zeta$\,  supposed for simplicity to be
isotropic:
\begin{eqnarray}
&G= &g_{cc}\nabla c^2+g_{\eta\eta}\nabla\eta^2+
g_{\zeta\zeta}\nabla\zeta^2\nonumber\\
&&+2g_{c\eta}\nabla c\nabla\eta +2g_{c\zeta}\nabla
c\nabla\zeta+2g_{\eta\zeta}\nabla\eta\nabla\zeta, \label{G}
\end{eqnarray}
while coefficients \,$g_{ij}$\, in this form  are supposed to be
constant.

 For simplicity, we also assume that in our model
functional (\ref{F_GL-tot}) the dependence of function
\,$f(c,\eta,\zeta)$\, on parameters \,$\eta$\, and \,$\zeta$\, can
be described by polynomials with at most the fourth order. Then, for
relations (\ref{eta-zeta}) for parameters \,$\eta$\, and \,$\zeta$\,
in equilibrium homogeneous phases to hold at all temperatures and
concentrations, the function  \,$f(c,\eta,\zeta)$\, can be taken in
the following form (analogous to that proposed in \cite{Plapp-02} to
model the directional solidification processes):
\begin{equation}
f(c,\eta,\zeta)=\varphi(c)+\Phi_1 (c,\eta)+\Phi_2 (c',\zeta)
\label{f_GL}
\end{equation}
where \,$c'=(1-c)$,\, and \,$\varphi(c)$\, is the free energy per
one interstice in a disordered alloy described for simplicity by the
mean-field approximation:
\begin{equation}
\varphi(c)=T(c\ln c +c'\ln c')-V_0cc'/2. \label{varphi_id}
\end{equation}
Positive  values \,$V_0$\, used below correspond to the effective
repulsion between interstitial atoms which qualitatively correctly
reproduces the type of carbon-carbon interactions in the real
austenite \cite{VKh-2}). The \,$\Phi_1 (c,\eta)$\, and \,$\Phi_2
(c',\zeta)$\, functions are taken in the following form:
\begin{eqnarray}
&\Phi_1(c,\eta)=&\lambda_1\,
c\,(\eta^2/2-\eta^3/3)\nonumber\\
&&+A_1\,[\tau_1\eta^2/2-(\tau_1
+1)\eta^3/3+\eta^4/4],\nonumber\\
&\Phi_2(c',\zeta)=&\lambda_2\,
c'\,(\zeta^2/2-\zeta^3/3)\nonumber\\
&&+A_2\,[\tau_2\zeta^2/2 -(\tau_2 +1)\zeta^3/3+\zeta^4/4].
\label{Phi_c-eta}
\end{eqnarray}
Here \,$\tau_1$\, is \,$T/2T_{\rm A}$\, where \,$T_{\rm A}$\, is the
temperature of the A--D phase transition at $c=0$; \,$\tau_2$\, is
\,T$/2T_{\rm B}$\, where \,$T_{\rm B}$\, is the B--D phase
transition temperature at $c'=0$ (i. e., $c=1$), and
\,$\lambda_1$,\, $\lambda_2$,\, \,$A_1$\, and \,$A_2$\, are some
positive energy parameters.

Values of the structure parameters \,$\eta$\, and \,$\zeta$\ in
equilibrium homogeneous phases are found by minimization of
functional (\ref{f_GL}) with respect to \,$\eta$\, and \,$\zeta$\,
at the given concentration \,$c$.\, The minimization with respect to
\,$\eta$\, yields equation \,$\partial f/\partial\eta
 =\partial\Phi_1/\partial\eta =0$ with the function \,$\Phi_1$\, from
 (\ref{Phi_c-eta}), while the minimization with respect to
\,$\eta$\, yields an analogous equation with the derivative of
function \,$\Phi_2$:\, with respect to \,$\zeta$:\,
\begin{eqnarray}
&&\eta (1-\eta )(\tau_1 - \eta +\lambda_1 c/A_1)=0,\label{eta_equilib}\\
&&\zeta (1-\zeta )(\tau_2 - \zeta
+\lambda_2c'/A_2)=0.\label{zeta_equilib}
\end{eqnarray}
Choice of physically acceptable solutions of these equations at
different \,$(T,c)$\,  values will be explained for the equation
(\ref{eta_equilib}); treatment of Eq. (\ref{zeta_equilib}}) is
similar. In Fig. 6 we show the \,$\eta$\,-dependence of the
\,$\Phi_1$\, function in Eq. (\ref{Phi_c-eta}) for two physically
different situations:
\begin{eqnarray}
&&0<(\tau_1 +\lambda_1 c/A_1)<0.5,\label{tau-lambda-1}\\
&&0.5<(\tau_1 +\lambda_1 c/A_1)<1.\label{tau-lambda-2}
\end{eqnarray}

In both  cases, the \,$\Phi_1 (\eta)$\, function has minima
at\,$\eta=0$\, and \,$1$\, (and a maximum at a certain intermediate
value \,$\eta=\tau +\lambda c/A_1$).\, However, in the case
(\ref{tau-lambda-1}), the  right minimum is below the left minimum,
so that phase A with \,$\eta=1$\, is thermodynamically favorable,
while  in the case (\ref{tau-lambda-2}), the left minimum is below
the right minimum, and phase D  with \,$\eta=0$\, is
thermodynamically favorable. The line \,$(\tau_1 +\lambda_1
c/A_1)=0.5$\, separating these two regions in the \,$(c,T)$\, plane
 is shown as a dashed line in the left part of figures 5(a)--5(c).
It corresponds to the line of phase transitions between phases A and
D in the absence of phase separation, i. e., at a constant,
``frozen'' concentration  \,$c$.\, Similarly, the dashed line in the
right part of figures figures 5(a)--5(c) shows the line of phase
transitions between phases B and D at a ``frozen'' concentration
\,$c$.\,

If we take into account possible phase separation, that is, if we
minimize the total free energy (\ref{F_GL-tot}) also with respect to
the number of  Me and X atoms in each phase, we can write total
equilibrium equations for any two phases, 1 and 2, in the well-known
form:
\begin{eqnarray}
&& (\partial f/\partial\eta)_{1}=(\partial f/\partial\eta)_{2}
=0;\quad
(\partial f/\partial\zeta)_{1}=(\partial f/\partial\zeta)_{2} =0;\nonumber\\
&& (\partial  f/\partial c)_1=(\partial  f/\partial c)_2=\mu; \quad
f_1-\mu c_1=f_2-\mu c_2\label{phase-equlib}
\end{eqnarray}

where  \,$\mu$\, is the  chemical potential of carbon with respect
to iron \cite{VZhKh}.

Simulations of phase transformations described below were made
mainly for three thermodynamic models: for the symmetric model 1
from \cite{VS-08} with the following values of parameters in Eqs.
(\ref{G})-(\ref{Phi_c-eta}):
\begin{eqnarray}
&&\lambda_{1,2}=A_{1,2}=5T_A,\quad T_A=T_B,\quad  V_0=0,\nonumber\\
&&g_{\eta\eta,\zeta\zeta}=a^2T_A,\quad
g_{cc,c\eta,c\zeta,\eta\zeta}=0,\label{model-1}
\end{eqnarray}
and for two non-symmetric models, 2 and 3, with the following
parameter values:
\begin{eqnarray}
\hskip-5mm&&{\rm Model\  2}:\quad  \lambda_1=7.5T_A,\quad A_1=5T_A,\quad \lambda_2=14T_A, \nonumber\\
\hskip-5mm&&A_2=9.49T_A,\quad T_B=2T_A,\quad V_0=0,\quad g_{cc}=3a^2T_A,\nonumber\\
\hskip-5mm&& g_{\eta\eta}=0.05a^2T_A, \quad
g_{\zeta\zeta}=0.5a^2T_A, \quad
g_{c\eta,c\zeta,\eta\zeta}=0;\label{model-2}\\
\hskip-5mm&&{\rm Model\  3}:\quad  \lambda_1=24T_A,\quad A_1=8T_A,\nonumber\\
\hskip-5mm&& \lambda_2=23T_A,\quad A_2=30T_A,\quad T_B=3.8T_A\,\nonumber\\
\hskip-5mm&&V_0=T_A, \quad g_{cc,\eta\eta,\zeta\zeta}=2T_Aa^2;\quad
g_{c\eta,c\zeta,\eta\zeta}=0, \label{model-3}
\end{eqnarray}
where \,$a$\, is the FCC lattice constant. Equilibrium phase
diagrams for these models are shown in Fig. 5. Eutectoid values of
concentration and temperature, \,$c_e$\, and \,$T_e$,\, for these
models are: for model 1, \,$c_e=1/2$,\, \,$T_e=0.43\,T_A$;\, for
model 2, \,$c_e=1/3$,\, \,$T_e=0.54\,T_A$;\, and for model 3,:
\,$c_e=1/8$,\, \,$T_e=0.83\,T_A$.\,  Note that  in the phase diagram
for model 3 shown in frame 5(c), the left binodal for the phase
equilibrium A--B coincides with the \,$c=0$\, axis within accuracy
of drawing: solubility limits \,$c_s(T)$\, are of the order of
\,$10^{-3}$.\,

\begin{figure}
\caption {(a)  Phase diagram temperature \,$T$\, - concentration
\,$c$\, for symmetric model 1 described by Eqs. (\ref{model-1}).
Phases \,$A$,\, \,$B$\,  and \,$D$\, correspond to ferrite,
cementite and austenite; solid lines are two-phase equilibrium
curves. Left or right dashed line shows the stability limit of phase
D with respect to transition to phase A or phase B at the fixed
concentration $c$. Circle shows the \,$(T,c)$\, values for which
simulations of growth of colonies have been made. \,(b)\,: Same as
in (a) but for model 2 described by Eqs. (\ref{model-2}). \,(c)\,:
Same as in (a) but for model 3 described by Eqs. (\ref{model-3}).
 \label{phase-diagrams}}
\end{figure}
\begin{figure}
\caption { Dependence of  \,$\Phi_1/A_1$\, on \,$\eta$\, in Eq.
(\ref{Phi_c-eta}), at the temperature and concentration falling in
the range determined by inequalities (\ref{tau-lambda-1}) (solid
line), and inequalities
 (\ref{tau-lambda-2}) (dotted line).
 \label{Phi_eta}}
\end{figure}
% \vskip 3mm

\subsection{Kinetic model\label{kinetic-model}}

Diffusion of interstitial (carbon) atoms in the course of pearlite
transformations is described using the quasi-equilibrium kinetic
equation method described in Ref. \cite{Vaks-04}. In the case of
weakly inhomogeneous states under consideration, this equation for
local concentration \,$c({\bf r},t)$\, takes the continuum form
\cite{PV-03}:
\begin{equation}
\partial c/\partial t =-{\rm div}{\bf j};\quad
j_{\alpha}=-cc'\sum_{\beta}D_{\alpha\beta}\nabla_{\beta}\left(\delta
F/\delta c\right).\label{c_dot}
\end{equation}
Here \,$\alpha$\, and \,$\beta$\,  are Cartesian indices,
\,$D_{\alpha\beta}$\, is the diffusivity tensor, while function
\,$\delta F/\delta c=\delta F/\delta c({\bf r})$\, (having the
meaning of the local chemical potential of carbon atoms) is the
variational derivative of functional (\ref{F_GL-tot}) with respect
to local concentration \,$c({\bf r})$:
\begin{equation}
\delta F/\delta c({\bf r})=\partial f(c,\eta,\zeta)/\partial c
-2(g_{cc}\Delta c+g_{c\eta}\Delta\eta+g_{c\zeta}\Delta\zeta)
\label{delta_F}
\end{equation}
where $\Delta=\nabla^2$ is the Laplace operator. In describing the
diffusivity \,${\bf D}$\,  we take into account not only the usual
volume  diffusivity \,${\bf D}_v$\, but also possible surface
contributions \,${\bf D}_s$\, which can be important due to the
enhanced diffusion of carbon along incoherent interphase boundaries
\cite{Hillert-62,Whiting-00,Bokshtein-61}. Therefore, the
diffusivity includes both volume and interfacial terms and is
written as follows:
\begin{equation}
D_{\alpha\beta}=D_v^{\alpha\beta}+a^2\sum_{\gamma\delta}
\varepsilon_{\alpha\gamma}\varepsilon_{\beta\delta}
(D_s^{\eta}\nabla_{\gamma}\eta\nabla_{\delta}\eta+
D_s^{\zeta}\nabla_{\gamma}\zeta\nabla_{\delta}\zeta).
 \label{D_alpha-beta}
\end{equation}

Here the first term  describes the volume diffusion (being,
generally, different in different phases which in our model can be
described by the dependence of \,$D_v^{\alpha\beta}$\, on \,$\eta$\,
and \,$\zeta$).\, Terms with \,$D_s^{\eta}$\, and \,$D_s^{\zeta}$\,
describe the surface diffusion of carbon atoms along interfaces
austenite-ferrite and austenite-cementite, respectively. Below we
discuss only growth of plane pearlite lamellas lying within
\,$(y,z)$\, plane. For these 2D problems, Cartesian indices
\,$\alpha,\beta,\gamma,\delta$\, in the second sum in
(\ref{D_alpha-beta}) are \,$x$\, or \,$y$,\, while
\,$\varepsilon_{\alpha\beta}= -\varepsilon_{\beta\alpha}$\, is the
unit antisymmetric tensor with just two non-zero components:
\,$\varepsilon_{xy}$\,=1 и \,$\varepsilon_{yx}$\,=(-1).

Temporal evolution of structure parameters \,${\eta}$\, and
\,${\zeta}$\, is described by the time-dependent Ginzburg-Landau
equations \cite{PV-03} with two phenomenological relaxation
parameters, \,$\gamma_{\eta}$\, and \,$\gamma_{\zeta}$:
\begin{eqnarray}
\hskip-10mm&\partial\eta/\partial t=&-\gamma_{\eta}\delta F/\delta
-\gamma_{\eta}\,[\partial \Phi_1 (c,\eta)/\partial\eta
\eta({\bf r})\nonumber\\
&&-2(g_{c\eta}\Delta c+g_{\eta\eta}\Delta\eta+g_{\eta\zeta}\Delta\zeta)];\nonumber\\
\hskip-10mm&\partial \zeta/\partial t =&-\gamma_{\zeta}\delta
F/\delta \zeta({\bf r})=-\gamma_{\zeta}\,[\partial \Phi_2
(c',\zeta)/\partial\zeta\nonumber\\
&&-2(g_{c\zeta}\Delta c+
g_{\eta\zeta}\Delta\eta+g_{\zeta\zeta}\Delta\zeta)].\label{eta-zeta_dot}
\end{eqnarray}

 As discussed in detail in Ref. \cite{VS-08}, ``phonon'' relaxation
times \,$\tau_{ph}\sim (\gamma_{\eta}^{-1},\gamma_{\zeta}^{-1})$\,
in Eqs. (\ref{eta-zeta_dot}) describe relaxation of crystal lattice
due to the anharmonic interactions of phonons, and they are by many
orders of magnitude shorter than the ``diffusional'' times
\,$\tau_d\sim Da^2$\, that describe relaxation of concentration
according to Eq. (\ref{c_dot}) being realized by the diffusional
jumps of carbon atoms between interstices. Therefore, one can expect
that in  the course of pearlite transformations under study,
structure parameters \,${\eta}$\, and \,${\zeta}$\, adiabatically
fast follow the slowly varying distribution of carbon concentration
 \,$c({\bf r},t)$,\, minimizing the free energy (\ref{F_GL-tot}) at each given
\,$c({\bf r},t)$.\, In particular, if the local concentration
\,$c=c({\bf r},t)$\, obeys inequality (\ref{tau-lambda-1}), the
\,$\eta({\bf r},t)$\, value should be close to unity (that is, the
lattice structure should be close to ferrite), while if inequality
(\ref{tau-lambda-2}) is obeyed, we should have: \,$\eta({\bf
r},t)\simeq 0$\,  (that is, the lattice structure should be close to
austenite), and analogously for \,$\zeta({\bf r},t)$\,. Deviations
from these ``uniform'' values of \,$\eta$\,  and \,$\zeta$\, (i. e.,
from their values of ``zero-order'' in non-uniformity) arise only
due to the presence of last, gradient terms in Eqs.
(\ref{eta-zeta_dot}) which in the Ginzburg-Landau approach used are
supposed to be small.

In our simulations, we describe this physical picture as follows.
The generalized diffusion equation (\ref{c_dot}) for the
concentration \,$c=c({\bf r},t)$\, is considered as the main one. In
the computations, it is replaced by its finite-difference analog and
is solved using standard iterative methods. However, after each step
of these iterations, the \,$c({\bf r},t)$\, values obtained are
divided into two groups for which either inequality
(\ref{tau-lambda-1}) or inequality (\ref{tau-lambda-2}) holds. For
points \,$\bf r$\, corresponding to the first group, we put
\,$\eta({\bf r},t)=1$,\, while for \,$\bf r$\, corresponding to the
second group, we put \,$\eta({\bf r},t)=0$.\, The analogous
procedure is made for the \,$\zeta({\bf r},t)$\, values. After that,
both parameters \,${\eta}$\, and \,${\zeta}$\, start to evolve with
time according to the ``phonon'' equations (\ref{eta-zeta_dot})
(again replaced by their finite-difference analogs) for the time
interval \hbox{\,$(t,t+\Delta t)$}\, at the fixed values \,$c({\bf
r},t)$.\, In these computations, we put for definiteness \,$
\gamma_{\zeta}=\gamma_{\eta}$,\, and the interval \,$\Delta
t\lesssim 0.5\gamma_{\eta}^{-1}$\, was found to be sufficient for
the full relaxation of \,$\eta$\, and \,$\zeta$\, to their
``quasi-equilibrium'' values \,$\eta[c({\bf r},t)]$\, and
\,$\zeta[c({\bf r},t)]$.\, Then these relaxed \,$\eta$\, and
\,$\zeta$\, values are put into Eq. (\ref{c_dot}) as the initial
values for the next iteration in \,$c({\bf r},t)$,\, and so on.

\section{Simulations of growth of eutectoid pearlite colonies}

\subsection{Methods of simulations\label{methods}}

Methods of simulations of the steady-state growth were basically the
same  as those in  Ref. \cite{VS-08}. We employed a simulation
volume \,$V_s=(L_x\times L_y \times L_z)$\, in the cubic lattice
with periodic boundary conditions along \,$y$\, and \,$z$\, axes.
The \,$y$\, axis was chosen along the colony growth direction, while
the period \,$L_z$ along \,$z$\, axis  was taken $a$,\, so that the
growth of lamellar colonies parallel to the  \,$yz$\, plane was
simulated. The simulation length  \,$L_x$  along \,$x$\, axis  was
taken as a half of the colony period \,$S$:\,  \,$L_x=S/2$,\, and
the mirror boundary conditions along \,$x$\, axis were used both at
\,$x$=0\, and  at \,$x$=$S/2$.\, Differential equations
(\ref{c_dot}) and (\ref{eta-zeta_dot}) were replaced by their
finite-difference analogs  with the  space-step \,$L_s$\, and the
time-step \,$t_s$,\, and these equations were solved by the standard
Runge-Kutta method. Values \,$L_s=a$\, and \,$t_s=10^{-3}\tau_d$\,
were usually employed where \,$\tau_d$\, has the meaning of a
characteristic time of diffusional jumps on the distance  \,$a$,\,
while the relations of this \,$\tau_d$\, to diffusivities \,$D_v$\,
or \,$D_s$\, in Eq. (\ref{D_alpha-beta}) are indicated below. The
initial distribution of parameters \,$c({\bf r})$, \,$\eta({\bf
r})$\, and \,$\zeta({\bf r})$\, (illustrated by frames 7a, 8a  and
11a) was chosen close to that expected for the steady-state growth
as described in detail in Ref. \cite{VS-08}.

Our simulations were usually performed for relatively low
temperatures: \,$T\lesssim 0.5T_e$,\, while actually pearlite
transformations are realized at higher \,$T\gtrsim 0.9T_e$.\,
However, the colony periods  \,$S$\, for such ``realistic'' \,$T$\,
become rather large: \,$S\gtrsim 500a$,\, and simulations of
evolution of colonies for such  \,$S$\, become time-consuming. At
the same time, the main aim of this work  is elucidation of just
main mechanisms but not quantitative details of transformations,
while qualitative manifestations of these mechanisms seem to be
weakly sensitive to the temperature \,$T$\, values. In addition to
that, in studies of the most important problems, such as the
instability of growth of pearlite colonies via the volume diffusion
mechanism discussed in Sec. \ref{volume-diffusion}, we also made
checking simulations at higher \,$T\sim 0.8T_e$,\, and the results
did not significantly change. Therefore, the main results and
conclusions presented below seem to be realistic  and  reliable in
spite of all simplifications in both models and simulation
parameters used.

\subsection{Growth of colonies via the volume diffusion
of carbon   mechanism \label{volume-diffusion}}

\begin{figure}
\caption{ Evolution of eutectoid colonies via the volume diffusion
mechanism for the symmetric model 1 with the phase diagram shown in
Fig. 5(a) at  temperature \,$T=0.25T_A$,\, the period \,$S=32a$,\,
and the following values of the reduced time \,$t'=t/\tau_d$:\,
\,(a) 0, (b) 50, and (c) 400. The grey level linearly varies with
local concentration \,$c({\bf r})$\, between 0 and 1 from white to
black. \label{vol-growth-model_1-S_32}}
\end{figure}
% \vskip 3mm
%
\begin{figure}
\caption {Same as in Fig. 3 but at  \,$S=64a$.\,
\label{vol-growth-model_1-S_64}}
\end{figure}
% \vskip 3mm

In Figs. 7-10 we present some results of our simulations of growth
of eutectoid colonies (i. e. growth of colonies into austenite with
the carbon concentration \,$c_a$\, equal to the eutectoid one,
\,$c_e$)\, via the  volume diffusion of carbon mechanism. For these
simulations we suppose that the surface diffusivities in Eq.
(\ref{D_alpha-beta}) are absent: \,$D_s^{\eta}$=\,$D_s^{\zeta}$=0,\,
while the volume diffusivity is isotropic:
\,$D_v^{\alpha\beta}$=$D_v\delta_{\alpha\beta}$\, and it has the
same value \,$D_v=a^2/\tau_d$\, in all three phases under
consideration, austenite, ferrite and cementite. Various
thermodynamic models used for these simulations differ mainly by the
eutectoid concentration \,$c_e$\, value. Figs. 7 and 8 correspond to
the symmetrical model 1 with \,$c_e$=$1/2,$\, and the phase diagram
shown in Fig. 5(a). Fig. 9 corresponds to  a ``not strongly''
asymmetric model 2 with \,$c_e$=$1/3$\,  and the phase diagram shown
in Fig. 5(b). Finally, in Fig. 10 we show results for the more
realistic, strongly asymmetric model 3 with \,$c_e$=$1/8.$\,

The results presented in Figs. 7 and 8 (as well as other results for
the growth of colonies in symmetrical models discussed in detail in
Ref. \cite{VS-08}) basically agree with conclusions of
phenomenological theories \cite{Hillert-72,Schastlivtsev-06}. At the
same time, Fig. 9 shows that for asymmetric models,  this growth
becomes not quite stationary: it is accompanied by some oscillations
in the colony front structure, where the cementite and ferrite
lamellas leave behind each other in turn. However, for a ``not
strongly'' asymmetric model 2, this non-stationarity seems to be not
very important.

However, for strongly asymmetric models, such as our realistic model
3 with \,$c_e$=$1/8,$\, the analogous instability of the growth
leads to the impossibility of formation of a regular structure of
pearlite colonies via the volume diffusion mechanism. It is
illustrated by Fig. 10 which shows that in this case, already for
rather short evolution times: \,$t\lesssim 10\tau_d$,\, the ferrite
lamellas start to leave behind the cementite lamellas and to fuse
with each other. It leads to the isolation (``divorcing'') of
cementite lamellas from austenite and thus to the locking of their
further growth.

The similar locking (which in literature is sometimes called
``divorcing of pearlite'' \cite{Hillert-62}) had been observed in
our simulations for all models with realistic \,$c_e\simeq 1/8$.\,
This phenomenon was found to be insensitive to varying thermodynamic
and gradient parameters \,$\lambda_i$,\, \,$A_i$,\, \,$T_{A,B}$,\,
\,$V_0$\, and \,$g_{ik}$\, in Eqs. (\ref{model-3}), as well as
temperature \,$T$,\, and the diffusivity \,$D_v$\, in ferrite and
cementite.

Such instability of  growth of eutectoid colonies at small
\,$c_e\simeq 1/8$\, seems to be related to the most general relation
between the diffusion time \,$t_d$\, and the diffusion length
\,$l_d$\, for the volume diffusion mechanism: \,$t_d\sim
l_d^2/D_v$.\, As the distance \,$l_{d}$\, needed for carbon atoms to
diffuse through austenite between centers of the ferrite and the
cementite lamellas much exceeds the cementite lamella half-width
\,$l_c$:\, \,$l_d^2\sim 50\,l_c^2\gg l_c^2$,\, the diffusion time
needed for carbon atoms released from a growing ferrite lamella to
reach cementite lamellas much exceeds times of ``consumption'' of
carbon atoms from the space before front of these cementite
lamellas. An amount of this fast consumed carbon is much lower than
that needed for a noticeable growth of this highly concentrated
cementite lamella. Therefore, it grows much slower than adjacent
ferrite lamellas whose growth is realized via diffusion of released
carbon atoms mainly ``forward'', into the non-transformed austenite,
rather than sideways, to the narrow and low-mobile cementite
lamellas. Thus the ferrite transformation front leaves behind the
cementite transformation front. Therefore, the steady-state transfer
of carbon atoms between ferrite and cementite lamellas (supposed to
occur in the phenomenological treatments
\cite{Zener-45}-\cite{Hillert-72}) is actually not realized.

\begin{figure}
\caption{ Same as in Fig. 7 but for the weakly asymmetric model 2
with the phase diagram shown in Fig. 5(b) at temperature
\,$T=0.2T_A$,\, period \,$S=60a$,\, and the following \,$t'$:\, (a)
100, (b) 300, (c) 500, and (d) 1000.
\label{vol-growth-model_c_e_1/3}}
\end{figure}
% \vskip 3mm

%
\begin{figure}
\caption{Same as in Fig. 7 but for model 3
 with the phase diagram shown in Fig. 5(c) at temperature \,$T=0.4T_A$,\, period
\,$S=48a$,\, and following \,$t'$:\, \,(a) 3, (b) 10, and (c) 12.
\label{vol-growth-model_2}}
\end{figure}
% \vskip 3mm

%
\begin{figure}
\caption{Evolution of eutectoid colonies  for model 3 via the
interfacial diffusion of carbon mechanism described in the text at
temperature \,$T=0.4T_A$\, and different periods \,$S$.\, Upper row:
\,$S=44a$,\, while the values of \,$t'=t/\tau_d$\, are: \,(a) 0, (b)
500, и (c) 1000. Middle row: \,$S=48a$,\, while \,$t'$\, are: \,(d)
100, (e) 400, and (f) 700. Lower row: \,$S=52a$,\, while \,$t'$\,
are: (g) 100, (h) 500, and (i) 1000. \label{surf-growth-stable}}
\end{figure}
% \vskip 3mm

\begin{figure}
\caption{Same as in Fig. \ref{surf-growth-stable} but at \,$S=40a$\,
and the following \,$t'$:\, \,(a) 50, (b) 100, and (c) 120.
\label{surf-growth-S_40}}
\end{figure}
% \vskip 3mm

\begin{figure}
\caption{Same as in Fig. \ref{surf-growth-stable} but at \,$S=56a$\,
 and the following \,$t'$:\,  \,(a) 50, (b) 250, and (c) 280.
\label{surf-growth-S_56}}
\end{figure}
% \vskip 3mm

%
\begin{figure}
\caption{ \,(A):\, Ratio of the steady-state colony growth rate
\,$V$\, to its maximum value \,$V_{max}$\, versus the reduced colony
period \,$S/S(V_{max})$.\, Solid line:  our simulations illustrated
by Fig. 11 for which we found: \,$V_{max}$=$0.031a/\tau_d$.\, Dotted
and dashed lines: results of phenomenological treatments
\cite{Hillert-57,Hillert-72} for the volume and surface diffusion
mechanism, respectively. \,(B):\, Distribution of colony periods
\,$S$\, observed in experiments \cite{Mehl-56}.\label{V(S/S_max}}
\end{figure}

\subsection{Growth of colonies via the interfacial diffusion of
carbon mechanism \label{surface-diffusion}}

As the austenite-pearlite interphase boundary is basically
incoherent \cite{Hillert-62}, one can expect the surface diffusivity
along this boundary to much exceed the volume diffusivity
\,$D_v^a$\, of carbon in austenite, similarly to the grain boundary
diffusivity which, according to measurements by Bokshtein et al.
\cite{Bokshtein-61}, exceeds  this volume diffusivity by several
orders of magnitude.  We also note that the volume diffusivity in
ferrite at temperatures under consideration exceeds that in
austenite by 2-3 orders of magnitude  \cite{Blanter-62}. Keeping in
mind all that and employing also considerations of simplicity, in
simulations of growth of colonies via the interfacial diffusion
mechanism we supposed the values of effective interfacial
diffusivities, proportional to quantities \,$D_s^{\eta}$\, and
\,$D_s^{\zeta}$\, in Eq. (\ref{D_alpha-beta}), and  the volume
diffusivity in ferrite, \,$D_v^f$,\, to be similar, while the volume
diffusivities in austenite and cementite, \,$D_v^a$,\, and
\,$D_v^c$,\, to be negligibly small. Therefore, in these simulations
we put:
\begin{eqnarray}
&&D_s^{\eta}=D_s^{\zeta}=20a^2/\tau_d,\qquad
D_v^f=a^2/\tau_d,\nonumber\\
&&D_v^a=D_v^c=0,\label{D_surface-eff}
\end{eqnarray}
where we also take into account that interphase boundaries in our
model  have widths  \hbox{$w\sim$(4-5)$a$}, thus the structure
parameters  gradients \,$\nabla\eta$\, and \,$\nabla\zeta$\, in Eq.
(\ref{D_alpha-beta}) can be estimated as:
$|\nabla\eta|\sim|\nabla\zeta|\sim (0.2$-$0.25)/a$.

Some results our simulations for model (\ref{D_surface-eff}) are
presented in Figs. 11-14. Let us discuss these results, First, Fig.
11 shows that the stable steady-state growth of colonies via the
interfacial diffusion mechanism is possible, unlike that via the
volume diffusion mechanism discussed in Sec. \ref{volume-diffusion}.
Fig. 11 also shows that both the growth rate \,$V$\, and the
stationary front shape notably vary with the colony period \,$S$\,
which qualitatively agrees with conclusions of phenomenological
treatments \cite{Zener-45}--\cite{Hillert-72}. At the same time, our
microscopic approach reveals many kinetic features which are absent
in these treatments. In particular, Figs. \hbox{11-14} show that the
growth rate  \,$V$\, depends on the period \,$S$\, much stronger
than in the phenomenological treatments, and the interval of
possible  periods \,$S$\, is limited not only from below (by the
minimal value \,$S_0$\, which, according to Zener \cite{Zener-45},
is determined by the balance between the volume gain and the surface
loss of free energy under colony growth), but also from above, by
some maximum value \,$S_{max}$\, related to the development at
\,$S>S_{max}$\, of the pearlite divorcing processes illustrated by
Fig. 13; these processes are analogous to those shown in Fig. 10.

The resulting dependence \,$V(S)$\, (shown in Fig. \,14(A)\, by
solid line) turns out to be  much sharper than the analogous
phenomenological dependences  (shown in Fig. \,14(A)\, by dotted and
dashed line) for both the volume and the surface diffusion
mechanism. At the same time, this more sharp dependence \,$V(S)$\,
seems to better agree with the distribution of periods \,$S$\,
observed in experiments and illustrated by Fig. \,14(B).\,

Let us also note that our main conclusion that the growth of
pearlite colonies is determined by the surface rather than volume
diffusion mechanism agrees with a similar conclusion  made by
Whiting \cite{Whiting-00} basing on his analysis of experimental
data about the pearlite growth velocities.

\section{Model of formation of pearlite colonies near grain
boundaries of austenite}

As mentioned, possible mechanisms of formation of pearlite colonies
are widely discussed in the literature
\cite{Mehl-56,Hillert-62,Schastlivtsev-06}. However, these
discussions include usually either detailed phenomenological
description of observations of these complex processes
\cite{Hillert-62}, or just general considerations about such
mechanisms \cite{Mehl-56,Schastlivtsev-06}, with no attempts of
microscopic treatments or modeling of these processes. In this
section we discuss a simple theoretical model of formation of
pearlite colonies near grain boundaries of austenite based on the
assumption of a great enhancement of carbon diffusivity in this
region.  We show that in the simplest form described below, this
model can be applied only to strongly deformed materials but not to
the usual materials with relaxed grain boundaries. However, some of
results described below can also be useful for understanding of
similar processes near relaxed grain boundaries, in particular,
those shown in Figs. 2-4 of the present work.

\subsection{Kinetic model}

The thermodynamics of transformation will be described by our most
realistic model 3 with parameters given by Eqs. (\ref{model-3}). In
treatment of kinetics, it is convenient to separately consider the
formation of colonies near grain boundaries of austenite and their
subsequent growth inside the grain. For brevity, these two stages of
evolution will be referred to as the ``formation'' and the ``growth
inside the grain'' stages. In simulations of the formation stage we
made the following assumptions.

(A) Within a layer of a width \,$h$\, adjacent to a plane grain
boundary, the volume diffusivity of carbon much exceeds both volume
and surface diffusivities within the grain. In our modeling, we
describe it by the following relation generalizing Eq.
(\ref{D_alpha-beta}) to this non-uniform case:
\begin{eqnarray}
&& 0<y<h:\quad
D_{\alpha\beta}=D_b\delta_{\alpha\beta},\nonumber\\
&&h<y:\quad D_{\alpha\beta}=D_b\delta_{\alpha\beta}\{1-\exp
[(y-h)/l]\}
 \label{D_b^alpha-beta}
\end{eqnarray}
where parameter  \,$l$\,  characterizes the width of transition to
the inner part of grain.

(B) This boundary layer is enriched by carbon and has concentration
\,$c_b> c_e$\, while the concentration within the grain is
\,$c_e$.\,

(C) At the initial time \,$t=0$,\, there exist a plane lamella of
cementite (or cementite with adjacent ferrite) which has a length
\,$h_0$,\, width \,$w_0$,\, and is normal to the grain boundary,\,
as illustrated by frames (a) in Figs. 15-19.

\begin{figure}
\caption{ Formation of colonies for the model described by relations
(A)-(C) in the text for the case \,$h,h_0\gg a$\, at
\,$c_b=c_e=0.125$,\, the simulation length along  \,$x$\, axis equal
to \,$L_x$=$512a$,\, and the following values of the reduced time
\,$t'=tD_b/a^2$:\,  (a) 0, (b) 200, (c) 300, и (d) 1000.
\label{plane-formation-c_b_0.125}}
\end{figure}
% \vskip 3mm

%
\begin{figure}
\caption {Same as in Fig. 15 but at  \,$c_b=0.25$,\, \,$L_x=240a$,\,
and the following \,$t'$:\, (a) 0, (b) 50, (c) 100, и (d) 300.
\label{plane-formation-c_b_0.25}}
\end{figure}

These assumptions qualitatively agree with the available
experimental observations and theoretical considerations. In
particular, Bokshtein et al. \cite{Bokshtein-61} found the carbon
diffusivity near grain boundaries in ferrite at  \,$T=550^{\rm o}$\,
to exceed that within the grain by 3-4 orders of magnitude, and they
observed  a similar (though somewhat lower) enhancement of diffusion
near grain boundaries of austenite. Authors of a recent theoretical
work \cite{Kesarev-10} discussed acceleration of diffusion near
grain boundaries in  strongly deformed materials; they concluded
that the enhancement of diffusion in such materials should spread
for the significant distances from grain boundaries. The enhanced
concentration of  carbon and carbides in vicinities of grain
boundaries was noted by a number of authors
\cite{Mehl-56,Bokshtein-61}.  For the processes  of formation of
colonies discussed above in connection with Figs. 2 and 3, such
enhancement seems to occur on the ``cementite'' sides of grain
boundaries, and so on.

Methods of simulations of formation of colonies for the model
(A)-(C) were basically the same as those described in Sec. 4.
Differences concern only boundary conditions. For
``one-dimensional'' simulations shown in Figs. 15 and 16, we put no
boundary conditions along \,$x$\, axis, while for the simulations
shown in Figs. 17-20, the mirror symmetry with respect to both plane
\,$x=0$\, and plane \,$y=0$\,  was supposed.

After a relatively fast formation of a colony near grain boundary
discussed above, its further growth into the grain was supposed to
occur via the interfacial diffusion mechanism described in Sec. 4.3.
To simulate this process, we used the following model.

(D) The initial carbon concentration  \,$c({\bf r},t$=$0)$=
$c_0(y)$\, starts to gradually decrease to its volume value
\,$c_e$\, when the distance \,$y$\,  from the grain boundary exceeds
the width \,$y_0$\, of the layer enhanced by carbon:
\begin{eqnarray}
&&y<y_0: \ c_0=c_b;\nonumber\\
&& y>y_0: \ c_0=c_e+(c_b-c_e)\exp \,[(y-y_0)/l_c]. \label{c_0-y}
\end{eqnarray}
In our simulations we used such values of parameters:
\,$c_b=0.25$,\, \,$c_e=0.125$,\, \,$y_0=40a$,\, \,$l_c=10a$.\,

(E) Diffusivity for this process corresponds to the interfacial
diffusion mechanism described by Eqs. (\ref{D_alpha-beta}) and
(\ref{D_surface-eff}).

(F) As the initial state for this modeling, we used the distribution
of parameters \,$c({\bf r})$,\, \,$\eta({\bf r})$\, and
\,$\zeta({\bf r})$\, obtained in the end of simulation of formation
of colonies for \,$h=h_0=13a$\, shown in Fig. 19. This distribution
is presented in frame 19(d) for the interval of \,$x$\, between
center of the second cementite lamella (nearest to the initial one)
and center of the third  ferrite lamella, with mirror boundary
conditions with respect to both boundaries of this interval along
$x$-axis.

\subsection{Results of simulations of processes of formation
 of pearlite colonies}

\begin{figure}
\caption {Same as in Fig. 15 but at  \,$c_b=0.25$,\, \,$L_x=240a$,\,
and the following \,$t'$:\, (a) 0, (b) 50, (c) 100, and (d) 300.
\label{plane-formation-c_b_0.25}}
\end{figure}

\begin{figure}
\caption{Same as in Fig. 15 but at \,$c_b=0.25$,\, \,$l=4a$,\,
\,$w_0=6a$,\, \,$h=h_0=20a$,\, \,$L_x=600a$,\, and the following
\,$t'$:\, (a) 0, (b) 100, (c) 500, (d) 1000, (e) 1700, and (f) 2800.
\label{formation_h=20a}}
\end{figure}

\begin{figure}
\caption {Same as in Fig. 17 but at \,$h=h_0=15a$,\, \,$L_x=480a$,\,
and the following \,$t'$:\,  (a) 0, (b) 100, (c) 500, (d) 1000, (e)
1700, and (f) 2300. \label{formation_h=15a}}
\end{figure}

\begin{figure}
\caption {Same as in Fig. 17 but at \,$h=h_0=13a$,\, \,$L_x=300a$,\,
and the following \,$t'$:\, (a) 0, (b) 100, (c)  1000,  and (d)
1600. \label{formation_h=13a}}
\end{figure}

\begin{figure}
\caption {Same as in Fig. 17 but at \,$h=h_0=10a$,\, \,$L_x=192a$,\,
and the following \,$t'$:\, (a) 100, (b) 500, and (c)  1000.
\label{formation_h=10a}}
\end{figure}

\begin{figure}
\caption{Growth of new-formed colonies into the grain for the model
described by relations (D)-(F)  at\,$L_x=73a$\, \,$L_y=73a$,\, and
the following \,$t'=t/\tau_d$:\, (a) 100, (b) 700, b (c) 1000, wjere
\,$\tau_d$\, is the same as in Sec. 4.3.
\label{growth-of-formed-colonies}}
\end{figure}

Some results of simulations based on the above-described models are
shown in Figs. 15--21. Let us first discuss the results presented in
Figs. 15 and 16 which correspond to a limiting case of a ``very
thick'' layer of enhanced diffusion: \,$h,h_0\gg a$,\, when the
evolution becomes effectively one-dimensional. Our simulations
showed that in this case, the perfectly periodic pearlite structures
are formed practically at any width of initial lamellas  \,$w_0$\,
and any concentration \,$c_b$\, within the layer, as for this
geometry there are no ``losses'' of carbon atoms from the phase
transformation zone. For example, when a ferrite lamella grows into
austenite, ``excess'' carbon atoms in austenite released from
ferrite are accumulated before the front of this lamella until their
local concentration exceeds the ``critical'' value that corresponds
to the right dashed curve in the phase diagram of Fig. 5(c) (for the
given temperature ,$T$).\, Then a cementite lamella starts to form
via the ``adiabatic'' mechanism described in Sec. 3.2. Later on this
lamella thickens consuming the carbon atoms before its front, and so
on. Note that when concentration \,$c_b$ in the layer is equal to
the eutectoid one, $c_b=c_e$,\, the period of a self-organized
colony obtained in the simulation shown in Fig. 15:  \,$S_b\simeq
39a$,\, is very close to the minimal period  of the steady-state
growth shown in Fig. 14(A): \,$S_0\simeq 40a$.\,

Figs. 17-20 illustrate evolution of microstructure for more
realistic cases of a finite width \,$h$\, of a layer of enhanced
diffusion. We found that the type of this evolution is weakly
sensitive to varying parameters \,$l$,\, \,$w_0$\, and \,$h_0$\,
mentioned in points (A) and (C), but it sharply depends on the width
\,$h$\, of the layer of enhanced diffusion. In particular, at
\,$h=20a$\, (Fig. 17), in the pearlite colony formed via this
mechanism survive  only  cementite lamellas  formed  ``in one'', see
frame 17(f). Values \,$h\sim (13$-$15)a$\, (Figs. 18 and 19) seem to
be ``optimal'' for formation of regular colonies, but their period
\,$S$\, sharply depends on the value of width \,$h$\, varying from
\,$S$$\simeq$$35a$\, at \,$h=15a$,\, to \,$S$$\simeq$$ 47a$\, at
\,$h=13a$.\, Finally, in too narrow layers: \,$h\lesssim 10a$\,
(Fig. 20), pearlite colonies do not form. Therefore, formation of
regular pearlite colonies via the model mechanism (A)-(C)  is
possible only if the width of the layer of enhanced diffusion
notably exceeds interatomic distances: \,$h\gtrsim 10a\gg a$.

If this condition is obeyed and a colony can be formed near grain
boundary via the mechanism (A)-(C), its further growth into the
grain can be described by the model (D)-(F) of the previous section.
To simulate this process we used the above-described formation model
with  \,$h$=$13a$\, for which the period of new-formed colonies
observed in Fig. 19: \,$S\simeq 47a$,\, is close to that
corresponding to the maximum steady-state growth rate and shown in
Fig. 14(A): \,$S(V_{max})=48a$\,. In Fig. 21 we present some results
of this simulation. One sees that the colony steadily grows with the
rate \,$V\sim V_{max}$\, which  seems to be natural for the model
used with \,$S\simeq S(V_{max})$.\,

Now let us discuss a possible relation of a simple pearlite
formation mechanism described in this section to reality. As
mentioned, such mechanism can be effective only if the enhanced
diffusion layer is sufficiently wide: \,$h\gtrsim (13-15)a$.\, For
the usual, relaxed grain boundaries such widths seem to be too high.
However, in materials subjected to an intense plastic strain, the
enhanced diffusion regions, according to theoretical estimates
\cite{Kesarev-10}, should significantly broaden, and the inequality
mentioned can be realized. In this connection, it can be interesting
to compare our figures 18 and 19 to some experimental data about
formation of pearlite colonies in strongly deformed steels obtained
by Tushinsky et al. \cite{Tushinsky-93} and presented in Figs. 22
and 23. Tushinsky et al. believe that these micrographs show
formation of pearlite colonies on subgrains of austenite arisen due
to the intense ``thermoplastic hardening'' of this steel. The
morphology of some colonies seen in these figures, in particular,
those positioned to the left of the center of Fig. 22, seems to be
rather similar to that shown in Figs. 18 and 19.

In the usual, not deformed steels, the pearlite colonies seem to be
formed via more complex, many-stage processes discussed in Sec. 2
and Ref. \cite{Hillert-62}. However, one can believe that the
peculiar features of transformations with strongly inhomogeneous and
anisotropic diffusivity described above can also be manifested in
these, many-stage paths of formation of pearlite colonies.

\section{Conclusions}

\begin{figure}
\caption{ Role of subgrains of austenite in formation of subcolonies
of pearlite (interrupted pearlite transformation) in a plastically
deformed steel. Symbol ``g.b.'' means grain boundary of austenite,
``s.b.'' means subgrain of austenite, and dark regions correspond
mainly to cementite  (photo 34 in \cite{Tushinsky-93}).
\label{Tushinsky-picture-1}}
\end{figure}

\begin{figure}
\caption{Formation of lamellas from initial globulae of cementite on
a subgrain of austenite  (photo 35 in \cite{Tushinsky-93}).
\label{Tushinsky-picture-2}}
\end{figure}

Let us summarize the main results of this work.

1. The microstructure of pearlite colonies in both non-deformed and
plastically strained eutectoid steels has been investigated using
the optical and the scanning electron microscopy methods. The
results obtained enable us to make some new conclusions about
mechanisms of formation of pearlite colonies, in particular, about
their many-step character and about differences of these mechanisms
for the processes of formation of colonies near grain boundaries in
eutectoid steels and near other lattice defects in non-eutectoid
steels.

2. A simple model for theoretical studies of pearlite
transformations is proposed. In spite of its simplicity, it seems to
reflect the most significant features of thermodynamics and kinetics
of phase transformations between austenite, ferrite and cementite.

3. Simulations of growth of pearlite colonies  based on this model
showed that the volume diffusion of carbon mechanism supposed in the
most of existing theories leads to the instability of the
steady-state growth of colonies at any parameters of the model and
any temperatures. A more adequate theory of growth based on the
mechanism of interfacial diffusion of carbon is suggested. The
kinetic characteristics of growth obtained in this theory differ
notably from those obtained in the standard phenomenological models
but they seem to better agree with available experimental data.

4. A model of formation of pearlite colonies based on the assumption
of a strong enhancement of carbon diffusion near grain boundaries of
austenite has been suggested. The results of simulations of
processes of formation of pearlite colonies based on this model seem
to qualitatively agree with the available data for plastically
deformed steels. Further generalizations of this model can help to
understand more complex many-stage processes observed under
formation of pearlite colonies in non-deformed steels.

\
\section*{ACKNOWLEDGMENTS}

The authors are much indebted to Profs. M. Hillert and V. V. Popov
for critical comments and valuable advices; to M. K. Zaluletdinov,
for participation in performing experiments,  and to Yu. N.
Gornostyrev,  V. N. Degtyarev  and. P. A. Korzhavy, for numerous
valuable discussions. The work was supported by  the Russian Fund of
Basic Research (grant No. 06-02-16476); by the fund for support of
leading scientific schools of Russia  (grant No. NS-3004.2008.2);
and by the program of Russian university scientific potential
development (grant  No. 2.1.1/4540).

\end{document}